\documentclass{kluwer}    
\usepackage[dvips]{graphicx}
\newdisplay{guess}{Conjecture}

\begin{document}                                                                                   
\begin{article}
\begin{opening}         
\title{A Grid of Synthetic Stellar UV Fluxes}
 
\author{Lino H. \surname{Rodriguez-Merino}}
\author{Miguel \surname{Chavez}}
\runningauthor{L. Rodriguez-Merino}
\institute{Instituto Nacional de Astrofisica, Optica y Electronica A.P. 51 y 216, \\
72000 Puebla, Mexico}

\author{Alberto \surname{Buzzoni}}
\institute{Telescopio Nazionale Galileo, A.P. 565, 38700 Santa Cruz de la Palma,
Spain, and \\
Osservatorio Astronomico di Brera, Merate, Italy}

\author{Emanuele \surname{Bertone}}
\institute{Osservatorio Astronomico di Brera, via Bianchi 46, 23807
Merate (Lc), Italy}


\begin{abstract}

We present preliminary results of a large project aimed at creating an extended 
theoretical and observational database of stellar spectra in the ultraviolet wavelength
range. This library will consist of IUE spectra at low and high resolution,
and a set of LTE and NLTE theoretical fluxes. A first grid of 50 model fluxes 
with solar metallicity, in the wavelength interval 1000$\div 4400$ \AA, is reported here. 
Calculations are based on the Kurucz (1993) SYNTHE code. The models span effective 
temperatures between 10\,000 K and 50\,000 K, and a surface gravity in the range 
$2.5 \leq \log g \leq 5.0$ dex.
\end{abstract}

\end{opening}           

\section{Introduction}

Over the past two decades, most of the work dealing with the study of stellar 
aggregates by using synthesis techniques has concentrated in the optical and near IR
wavelength interval. Detailed analyses have been carried out to 
investigate the usefulness of many absorption features as indicators of age 
and/or global chemical composition of stellar populations \cite{Buzzoni95}. 
Although in the UV there are several comprehensive 
works \cite{Fanelli,TBrown}, this spectral window has not yet been fully exploited 
and represents a promising tool for complementing our current knowledge. Moreover, 
at intermediate 
and high redshift the UV spectra of galaxies are detected in the optical
and therefore a characterization of UV spectral features is mandatory for the
interpretation of data collected with large telescopes.

\section{Synthetic Spectra}

Relying on the Kurucz (1993) SYNTHE code for model atmospheres, we computed 
a preliminary grid of 50 high-resolution ($\lambda/\Delta\lambda=$50\,000) synthetic spectra
in the wavelength interval from 1000 \AA\ to 4400 \AA. Spectra cover an effective 
temperature range between 10\,000~K and 50\,000~K at step of 5000~K, a surface 
gravity from $\log g = 2.5$ to 5.0 at step of 0.5 dex, and adopt a solar chemical 
composition. In Fig.~1 we show the behavior of metallic  absorption features
(mainly Fe) in the interval $1530\div 1570$ \AA\ vs.\ effective temperature 
and gravity. 
We plan to define a set of spectroscopic indices that measure the intensity 
of absorption lines and calibrate them  according to the atmosphere fundamental
parameters. This will eventually result in a set of fitting functions for use in
population synthesis studies.

\begin{figure}
\begin{center}
\includegraphics[width=\hsize]{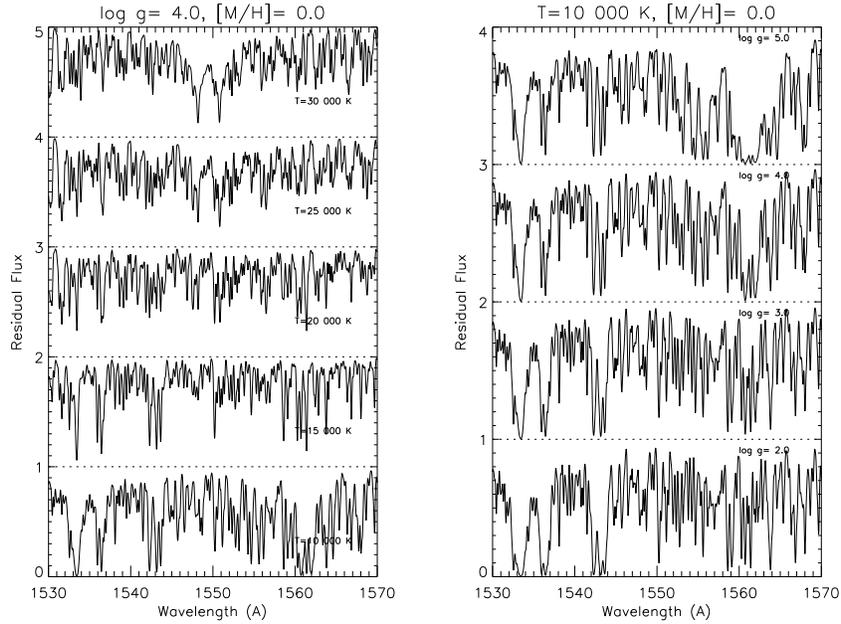}
\caption[]{Dependence of prominent UV absorption features, in the
1530$\div$1570 \AA\ wavelength range, vs.\ T$_{eff}$ 
and $\log g$. Theoretical spectra, from the Kurucz (1993) synthesis code,
are shown at the IUE resolution of 0.2 \AA\ FWHM.}
\end{center}
\end{figure}

\end{article}
\end{document}